\documentclass{article}
\usepackage{spconf,amsmath,graphicx,hyperref}
\def\BibTeX{{\rm B\kern-.05em{\sc i\kern-.025em b}\kern-.08em
    T\kern-.1667em\lower.7ex\hbox{E}\kern-.125emX}}
\usepackage{balance}
\usepackage{latexsym,mathrsfs}
\usepackage{amsmath,amssymb} 
\usepackage{amsthm,enumerate,verbatim}
\usepackage{amsfonts}
\usepackage{graphicx}
\usepackage[dvipsnames]{xcolor}
\usepackage{url}
\usepackage{pgfplots}
\pgfplotsset{compat=1.18}
\usepgfplotslibrary{statistics}
\usepackage{subcaption}
\usepackage{array}
\usepackage{diagbox}
\usepackage{cleveref}
\usepackage{enumitem}
\usepgfplotslibrary{fillbetween}
\usepgfplotslibrary{groupplots}
\usepackage{booktabs}
\usepackage{multirow}
\pgfplotsset{compat=1.17}
\usepackage{cite}
\usepackage{stmaryrd}
\usepackage{algorithm}    % provides the float environment + \caption
\usepackage{algorithmic}  % provides \REQUIRE, \ENSURE, \STATE, \WHILE, \ENDWHILE
\newtheorem*{definition*}{Definition}

\usetikzlibrary{decorations.pathreplacing, calligraphy}
\DeclareMathOperator*{\argmin}{argmin}

\Crefname{figure}{Fig.}{Figs.}
\crefname{algorithm}{Alg.}{Algs.}
\Crefname{algorithm}{Algorithm}{Algorithms}

\definecolor{myblue}{rgb}{0.0, 0.5, 1}

\definecolor{brightpink}{rgb}{1.0, 0.0, 0.5}

\definecolor{evrimteal}{rgb}{0.0, 0.55, 0.55}

\definecolor{roelviolet}{rgb}{0.45, 0.0, 0.75}

\newcommand{\R}{\mathbb{R}}
\newcommand{\email}[1]{\href{mailto:#1}{\nolinkurl{#1}}}

\newcommand{\Sec}[1]{\hyperref[sec:#1]{\S\ref*{sec:#1}}} %section
\newcommand{\Section}[1]{\hyperref[sec:#1]{Section~\ref*{sec:#1}}} %section
\newcommand{\AppFull}[1]{\hyperref[sec:#1]{Appendix~\ref*{sec:#1}}} %appendix
\newcommand{\Eqn}[1]{\hyperref[eq:#1]{(\ref*{eq:#1})}} %equation
\newcommand{\Fig}[1]{\hyperref[fig:#1]{Figure~\ref*{fig:#1}}} %figure
\newcommand{\Tab}[1]{\hyperref[tab:#1]{Table~\ref*{tab:#1}}} %table
\newcommand{\Thm}[1]{\hyperref[thm:#1]{Theorem~\ref*{thm:#1}}} %theorem
\newcommand{\Cor}[1]{\hyperref[cor:#1]{Corollary~\ref*{cor:#1}}} %corollary
\newcommand{\Alg}[1]{\hyperref[alg:#1]{Algorithm~\ref*{alg:#1}}} %algorithm
\newcommand{\Def}[1]{\hyperref[def:#1]{Definition~\ref*{def:#1}}} %definition

\newcommand{\Tra}{^{{\sf T}}} % transpose
\newcommand{\Inv}{^{-1}} % inverse
\newcommand{\V}[1]{{\bm{\mathbf{\MakeLowercase{#1}}}}} % vector
\newcommand{\Oprod}{\circ} % outer product

\newcommand{\M}[1]{{\bm{\mathbf{\MakeUppercase{#1}}}}} % matrix
\newcommand{\Khat}{\odot} %Khatri-Rao
\newcommand{\Hada}{\ast} %Hadamard

\newcommand{\T}[1]{\boldsymbol{\mathscr{\MakeUppercase{#1}}}} %tensor
\newcommand{\That}[1]{\boldsymbol{\hat \mathscr{\MakeUppercase{#1}}}} %tensor
\usepackage[mathscr]{eucal}
\usepackage{bm}          % for \bm
\usepackage{mathrsfs}    % for \mathscr
\title{Nonlinear Tensor Decomposition for Pattern Discovery}

\name{Atharva Awari$^{\star}$, Geert Roelof van der Ploeg$^{\dagger}$, Arnaud Vandaele$^{\star}$, Nicolas Gillis$^{\star}$, Evrim Acar$^{\dagger}$}
\address{$^{\star}$Department of Mathematics and Operational Research, Universit\'{e} de Mons (UMONS), Belgium\\
         $^{\dagger}$Simula Metropolitan Center for Digital Engineering, Oslo, Norway\\
         \small\texttt{\{atharvaabhijit.awari, arnaud.vandaele, nicolas.gillis\}@umons.ac.be}, \texttt{\{roel, evrim\}@simula.no}}

\begin{document}

\maketitle

\begin{abstract}
%The CANDECOMP/PARAFAC (CP) decomposition is widely used for interpretable pattern discovery \EA{from multiway datasets (also referred to as higher-order tensors}. When the data is clipped or saturated, however, the linear CP model becomes unreliable. Several remedies exist, but which one works depends on properties of the missingness that the practitioner must first diagnose \EA{Several remedies exist such as treating the clipped entries as missing or imputation; however, both fall short under various settings. In this paper,} we propose a nonlinear CP model (NCP) for pattern discovery in this setting, extending ideas from nonlinear matrix decompositions to tensors, and develop an ADMM algorithm for solving it. The ADMM framework is modular and flexible, and a variety of nonlinearities can be accommodated with minimal changes. On synthetic tensors with known factors, and on simulated metabolomics data where clipping usually arises from instrument detection limits, NCP recovers the underlying factors more faithfully than fitting CP to the observed entries alone or to threshold-imputed data. The advantage of NCP grows as clipping becomes more severe and without requiring the missingness pattern to be diagnosed in advance. \ngc{not 100\% clear what it means to 'diagnose the missingness pattern'  (also above). Be more specific.}
The CANDECOMP/PARAFAC (CP) decomposition is widely used for revealing the underlying patterns from multiway data (also referred to as a higher-order tensor). When the data is clipped or saturated, however, the linear CP model becomes unreliable. Several remedies exist such as treating the clipped entries as missing or imputation; however, both fall short in different scenarios. We propose a nonlinear CP model (NCP) for pattern discovery in this setting, extending ideas from nonlinear matrix decompositions to tensors, and solve it using a flexible ADMM (Alternating Direction Method of Multipliers)-based framework that can accommodate a variety of nonlinearities. Using synthetic data with known factors and simulated metabolomics data where clipping arises from instrument detection limits, we demonstrate that NCP recovers the underlying factors more accurately than CP fitted to only the observed entries or threshold-imputed data. %The advantage of NCP grows as clipping becomes more severe and without requiring the missingness pattern to be diagnosed in advance. \ngc{not 100\% clear what it means to 'diagnose the missingness pattern'  (also above). Be more specific.}

\end{abstract}

\begin{keywords}
CP, nonlinear decomposition, ADMM, clipping, saturation, pattern discovery, metabolomics.
\end{keywords}

% ══════════════════════════════════════════════════════════════════════════
\section{Introduction}
\label{sec:intro}
% ══════════════════════════════════════════════════════════════════════════

Tensor factorizations are a central tool for interpretable pattern discovery from multiway data \cite{morup2025tensor,kolda2009}. The most common model, CANDECOMP/PARAFAC (CP)~\cite{carroll1970analysis,harshman1970foundations}, represents a third-order tensor $\T{X} \in \R^{I \times J \times K}$ using $R$ components as
%, which represents a higher-order tensor as the sum of minimum number of rank-one tensors
%An $R$-component CP model represents a third-order tensor $\T{X} \in \R^{I \times J \times K}$ as
\begin{equation}
  \T X \;\approx\; \sum_{r=1}^{R} \V a_r \Oprod \V b_r \Oprod \V c_r
  \;=\; \llbracket \M A, \M B, \M C \rrbracket,
  \label{eq:cp}
\end{equation}
with factor matrices $\M A \in \R^{I \times R}$, $\M B \in \R^{J \times R}$, $\M C \in \R^{K \times R}$ and $\Oprod$ the outer product (see~\Cref{fig:nlcp}). We consider third-order tensors, but extension to higher orders is immediate. Unlike matrix factorizations, CP is essentially unique under mild conditions~\cite{kruskal1977}, facilitating interpretation of extracted components.
%which is what makes its components directly interpretable.

The CP model~\eqref{eq:cp} is fitted directly to observed data, and is therefore appropriate when the observations themselves are well approximated by a low-rank model. Deviations due to noise are handled effectively. However, when data is \emph{clipped}, information beyond a known threshold is lost. Values outside the measurable range are either recorded at the threshold itself or set to missing. %, in which case we only know that the threshold was crossed. 
In either case, the observation carries only the information that the true value lies beyond the threshold, so the data may no longer be low rank. Fitting~\eqref{eq:cp} directly then forces the CP factors to account for the underlying patterns as well as the clipping process. 
For instance, in audio clipping, samples whose amplitude exceeds the recording range are truncated, and the clipped value then acts as a lower bound on the true magnitude~\cite{kitic2014audio}. % which declipping methods exploit to recover the signa 
%In metabolomics, concentrations below the instrument's limit of detection (LOD) are unreported, so the LOD upper-bounds each censored value~\cite{wei2018missing}. 
In metabolomics, concentrations below the instrument limit of
detection (LOD) may go unreported, so the LOD upper-bounds each censored
value~\cite{wei2018missing}. %and concentrations exceeding the instrument's range saturate at its maximum, which then lower-bounds the true value~\cite{QuHi26}. 
In both cases the threshold is informative about what was lost. Affected entries are typically either treated as missing, discarding the threshold entirely, or imputed at the threshold~\cite{QuHi26}, treating it as the true value — a poor approximation when the underlying quantity lies far beyond it. Neither uses the threshold for what it is: a bound on the unobserved value.
%In this paper, we propose a \emph{nonlinear} CP model,

Motivated by the growing use of CP models in application areas where clipping is common, such as omics data analysis~\cite{yan2024characterizing, martino2021context}, we propose a \emph{nonlinear} CP model,
\begin{equation}
  \T X \;\approx\; f\bigl(\llbracket \M A, \M B, \M C \rrbracket\bigr),
  \label{eq:nlcp_intro}
\end{equation}
where an element-wise nonlinearity $f$ maps every value beyond the threshold onto it (see~\Cref{fig:nlcp}). The low-rank part $\llbracket \M A, \M B, \M C \rrbracket$ is then free to describe the underlying signal, while $f$ accounts for the clipping,  using $f(\cdot) = \min \big(b, \max ( \cdot, a)\big)$ for some interval $[a,b]$.  Such nonlinear decompositions were introduced in the matrix setting by Saul~\cite{saul2022nonlinear}, and extended %by Awari et al.
in~\cite{awari2025alternating} to a broader family of nonlinear models. Recent work has extended nonlinear decompositions to tensors using the ReLU nonlinearity to compress sparse nonnegative data~\cite{wang2026relu,wang2025simple}. However, these methods solve a latent-variable formulation that is not equivalent to the ReLU problem~\cite{gillis2026extrapolated}.
%The tensor case has been considered recently, only for the ReLU nonlinearity~\cite{wang2026relu, wang2025simple}, with the aim of compressing sparse nonnegative data, where the proposed approach solves a latent-variable formulation that is not equivalent to the ReLU problem itself~\cite{gillis2026extrapolated}, and therefore does not tackle it directly. %\ngc{I think we can remove this last sentence, too much detail, not very useful for our purpose.} 
We instead address the nonlinear CP problem directly and for a wide class of nonlinearities. 
Neural tensor models~\cite{chen2020neural} and neural additive tensor decomposition~\cite{ahn2024neural} also incorporate nonlinearity; however, they learn nonlinearities to increase the expressive power of tensor factorizations for prediction on sparse data.
%Another related line of research are neural tensor models~\cite{chen2020neural} and neural additive tensor decomposition~\cite{ahn2024neural} that introduce nonlinearity into tensor factorization for sparse data. %the latter explicitly to retain interpretability of the extracted components. 
%In both, however, the nonlinearity is a learned model parameter, introduced to increase expressive power on prediction tasks. 
In our case, the nonlinearity is fixed by the application rather than learned, allowing the threshold to be used as a bound. 
Previous work on nonlinear tensor methods has so far not addressed the pattern discovery problem. %(when part of the data is known only up to a threshold.)
%Pattern discovery in this setting, where part of the data is known only up to a threshold, has not been addressed.

Here, we formulate the nonlinear CP problem
in~\eqref{eq:nlcp_intro} and develop an alternating direction method of multipliers (ADMM) algorithm to solve it directly,
building on the framework recently proposed for nonlinear matrix
decompositions (NMDs)~\cite{awari2025alternating}. %The resulting algorithm is modular, accommodating different nonlinearities and constraints on the factors, also capable of incorporating loss functions other than the Frobenius norm. 
We instantiate it with the  MinMax nonlinearity and nonnegativity constraints on the factor matrices where required. On synthetic data with known ground-truth and simulated metabolomics data, we show
that the proposed nonlinear CP (NCP) model recovers the underlying factors more faithfully than treating the clipped entries as missing or imputing them at the threshold.
%, and that it does so without requiring the structure of the missingness to be known in advance.
% ══════════════════════════════════════════════════════════════════════════
\section{Proposed Method: Nonlinear CP--ADMM}

%We now introduce our proposed NCP model, and then provide an ADMM-based algorithm to estimate its parameters. 

%\paragraph*{A. The nonlinear CP (NCP) model.}
%\label{ssec:nonlinear_cp} 

\begin{figure}[t]
\centering
\resizebox{0.9\columnwidth}{!}{%
\begin{tikzpicture}[
    bar/.style   = {draw=black, line width=0.35pt},
    modeA/.style = {bar, fill=green!55!black},
    modeB/.style = {bar, fill=violet!60},
    modeC/.style = {bar, fill=orange!85!black},
]

% ---------- geometry ----------
\def\ch{0.95}      % height of the vertical (mode-1) bar
\def\cw{0.16}      % width  of the vertical bar
\def\bw{0.62}      % width  of the horizontal (mode-2) bar
\def\bh{0.14}      % height of the horizontal bar
\def\sk{0.20}      % skew   of the slanted (mode-3) bar

% ---------- one rank-one component ----------
\newcommand{\component}[2]{%
  % #1 = x offset, #2 = label
  \begin{scope}[shift={(#1,0)}]
    % mode-1 (vertical, green)
    \fill[modeA] (0,0) rectangle (\cw,\ch);
    % mode-2 (horizontal, violet)
    \fill[modeB] (0,\ch) rectangle (\bw,{\ch+\bh});
    % mode-3 (slanted, orange)
    \fill[modeC] (0,{\ch+\bh}) -- ({\sk},{\ch+\bh+\sk})
                 -- ({\sk+\cw},{\ch+\bh+\sk}) -- (\cw,{\ch+\bh}) -- cycle;
    \node[font=\tiny] at ({\bw/2},{\ch+\bh+\sk+0.18}) {#2};
  \end{scope}
}

% ---------- data tensor ----------
\begin{scope}[shift={(-0.15,0.20)}]
  \fill[gray!25, draw=black, line width=0.35pt]
      (0,0) rectangle (0.75,0.75);
  \fill[gray!15, draw=black, line width=0.35pt]
      (0,0.75) -- (0.22,0.97) -- (0.97,0.97) -- (0.75,0.75) -- cycle;
  \fill[gray!35, draw=black, line width=0.35pt]
      (0.75,0) -- (0.97,0.22) -- (0.97,0.97) -- (0.75,0.75) -- cycle;
\end{scope}
\node[font=\small] at (0.35,-0.42) {$\T X$};

% ---------- approx sign ----------
\node[font=\small] at (1.25,0.55) {$\approx$};

% ---------- f ( ----------
\node[font=\small] at (1.70,0.55) {$f$};
\node[font=\large] at (1.95,0.55) {$\Biggl($};

% ---------- components ----------
\component{2.25}{Comp.\ 1}
\node[font=\small] at (3.25,0.55) {$+$};
\component{3.55}{Comp.\ 2}
\node[font=\small] at (4.55,0.55) {$+\;\cdots\;+$};
\component{5.55}{Comp.\ $R$};

% ---------- ) ----------
\node[font=\large] at (6.55,0.55) {$\Biggr)$};

% ---------- brace: linear CP ----------
\draw[decorate, decoration={calligraphic brace, amplitude=3pt},
      line width=0.5pt, gray!70!black]
      (6.35,-0.30) -- (2.25,-0.30);
\node[font=\scriptsize, gray!40!black] at (4.30,-0.62) {linear CP};

\end{tikzpicture}%
}
\caption{The nonlinear CP model.}
\label{fig:nlcp}
\end{figure}
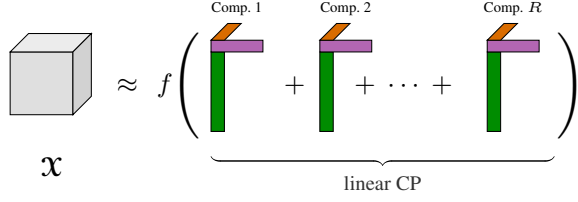

\noindent \textbf{A. The nonlinear CP (NCP) model.} 
Taking inspiration from the literature on NMDs~\cite{awari2025alternating}, we consider a NCP model (see \Cref{fig:nlcp}), where $f$ is an element-wise nonlinear function,
\begin{equation}
  \T X \;\approx\; f\bigl(\llbracket \M A, \M B, \M C \rrbracket\bigr),
  \quad
  x_{ijk} \approx f\Bigl(\textstyle\sum_{r=1}^{R} a_{ir}b_{jr}c_{kr}\Bigr).
  \label{eq:nlcp}
\end{equation}
Since $f$ is determined by the application and is known, it is not a
free parameter of the model. In this paper, we focus on the  \emph{MinMax} nonlinearity,
%\begin{equation*}
  $f(t) = \min\bigl(b,\, \max(a,\, t)\bigr)$,
  %\label{eq:minmax}
%\end{equation*}
which is used for two-sided clipping between known bounds $a$ and $b$. It covers one-sided clipping as a special case, e.g., the \emph{ReLU} function with $a = 0$ and $b = \infty$, with  $f(t) = \max(0,t)$.

%\paragraph*{B. ADMM algorithm.}
%\label{ssec:admm}  
\noindent \textbf{B. ADMM algorithm.}
Using least squares, fitting the NCP~\eqref{eq:nlcp} %to a third-order tensor $\T X$ 
requires to solve 
\begin{equation}
  \min_{(\M A, \M B, \M C) \,\in\, \mathcal{C}} \;
  \bigl\lVert \T X - f\bigl(\llbracket \M A, \M B, \M C \rrbracket\bigr)
  \bigr\rVert_F^2 ,
  \label{eq:nlcp_problem}
\end{equation}
where $\mathcal{C} = \mathcal{C}_A \times \mathcal{C}_B \times \mathcal{C}_C$ denotes the feasible set %\ngc{This is a rather strong abuse of notation: A,B,C do not have the same dimension, and could have different constraints... Maybe use $?} 
of the factor matrices, e.g.,  unconstrained, or the nonnegative orthant when the application calls for it~\cite{cichocki2009nonnegative}. Problem~\eqref{eq:nlcp_problem} is difficult to solve directly. Following the ADMM framework for NMDs~\cite{awari2025alternating}, we %simplify it by 
introduce an auxiliary tensor $\T T$: 
\begin{equation*}
  \min_{(\M A, \M B, \M C) \,\in\, \mathcal{C}, \; \T T} \;
  \lVert \T X - f(\T T) \rVert_F^2
  \quad\text{s.t.}\quad
  \T T = \llbracket \M A, \M B, \M C \rrbracket .
  \label{eq:admm_problem}
\end{equation*}
The nonlinearity now acts only on $\T T$, while the low-rank structure lives only in the constraint. With a scaled Lagrange multiplier tensor $\bm{\Lambda} \in \R^{I \times J \times K}$ and penalty $\rho > 0$, the
scaled augmented Lagrangian is
\begin{equation}
  L(\M A,\M B,\M C,\T T, \bm{\Lambda})
  = \lVert \T X - f(\T T) \rVert_F^2
  + \frac{\rho}{2}
    \bigl\lVert \T T - \llbracket \M A,\M B,\M C \rrbracket + \bm{\Lambda}
    \bigr\rVert_F^2 .
  \label{eq:aug_lag}
\end{equation}
ADMM alternates between minimizing~\eqref{eq:aug_lag} with respect to each primal block and updating the dual variable; see~\Cref{alg:admm_nlcp}. 
\begin{algorithm}[htb!]
\caption{ADMM for nonlinear CP decomposition}
\label{alg:admm_nlcp}
\begin{algorithmic}[1]
\renewcommand{\algorithmicrequire}{\textbf{Input:}}
\renewcommand{\algorithmicensure}{\textbf{Output:}}
\REQUIRE Tensor $\T X \in \R^{I \times J \times K}$, rank $R$,
nonlinearity $f$, feasible set $\mathcal{C} = \mathcal{C}_A \times \mathcal{C}_B \times \mathcal{C}_C$, penalty $\rho$,
max.\ iter $S$.
\ENSURE Factors $ (\M A, \M B, \M C) \in \mathcal{C}$ with
$\T X \approx f(\llbracket \M A,\M B,\M C \rrbracket)$.
\STATE \textbf{Initialization:} $(\M A, \M B, \M C) \in \mathcal{C}$,
$\T T \gets \T X$, $\bm{\Lambda} \gets 0$.
\WHILE{$s < S$ \textbf{ and } not converged}
   \STATE  $
    \M A \gets \argmin_{\M A \in \mathcal{C}_A}
    \bigl\lVert (\T T{+}\bm{\Lambda})_{(1)} - \M A (\M C \Khat \M B)\Tra
    \bigr\rVert_F^2.
    $ \vspace{0.1cm}
    
    \STATE $
    \M B \gets \argmin_{\M B \in \mathcal{C}_B}
    \bigl\lVert (\T T{+}\bm{\Lambda})_{(2)} - \M B (\M C \Khat \M A)\Tra
    \bigr\rVert_F^2.
    $  \vspace{0.1cm}
    \STATE
    $
    \M C \gets \argmin_{\M C \in \mathcal{C}_C}
    \bigl\lVert (\T T{+}\bm{\Lambda})_{(3)} - \M C (\M B \Khat \M A)\Tra
    \bigr\rVert_F^2.
    $  \vspace{0.1cm}
    
    \STATE \label{step:updateT} \textbf{Update $\T T$:} solve $IJK$ element-wise subproblems\\
    $
    t \gets \argmin_{t}\;
    \bigl(x_{ijk} - f(t)\bigr)^2
    + \tfrac{\rho}{2}(t + \lambda_{ijk} - \hat{x}_{ijk})^2 ,
    $
    where $\That{X} = \llbracket \M A,\M B,\M C \rrbracket$.
    \STATE 
    $\bm{\Lambda} \gets \bm{\Lambda} + \bigl(\T T - \llbracket \M A,\M B,\M C
    \rrbracket\bigr)$. 
\ENDWHILE
\end{algorithmic}
\end{algorithm} 
% \begin{subequations} \label{eq:admm_updates}
% \begin{align}
%   \M A^{s+1} &:= \argmin_{\M A \in \mathcal{C}}\;
%     L(\M A, \M B^{s}, \M C^{s}, \T T^{s}, \bm{\Lambda}^{s}), \label{eq:upd_A}\\
%   \M B^{s+1} &:= \argmin_{\M B \in \mathcal{C}}\;
%     L(\M A^{s+1}, \M B, \M C^{s}, \T T^{s}, \bm{\Lambda}^{s}), \label{eq:upd_B}\\
%   \M C^{s+1} &:= \argmin_{\M C \in \mathcal{C}}\;
%     L(\M A^{s+1}, \M B^{s+1}, \M C, \T T^{s}, \bm{\Lambda}^{s}), \label{eq:upd_C}\\
%   \T T^{s+1} &:= \argmin_{\T T}\;
%     L(\M A^{s+1}, \M B^{s+1}, \M C^{s+1}, \T T, \bm{\Lambda}^{s}), \label{eq:upd_T}\\
%   \bm{\Lambda}^{s+1} &:= \bm{\Lambda}^{s}
%     + \bigl(\T T^{s+1} - \llbracket \M A^{s+1},\M B^{s+1},\M C^{s+1}
%       \rrbracket\bigr). \label{eq:upd_L}
% \end{align}
% \end{subequations}
Crucially, $f$ appears only in the loss, $\lVert \T X - f(\T T)\rVert_F^2$, hence only the $\T T$-update (see step~\ref{step:updateT}) depends on the loss and nonlinearity. 
The factor and dual updates are independent of the choice of nonlinearity, and the constraint set $\mathcal{C}$ affects only the factor updates. This separation is what makes the framework modular, capable of accommodating different nonlinearities, constraints and loss functions.

\noindent\textbf{Factor updates.}
We derive the update for $\M A$; those for $\M B$ and $\M C$ follow
similarly. Dropping the loss term in \eqref{eq:aug_lag}, which does not depend on $\M A$, and unfolding
along the first mode, the $\M A$-update becomes
\begin{equation}
  \M A := \argmin_{\M A \in \mathcal{C}_A}\;
  \bigl\lVert (\T T + \bm{\Lambda})_{(1)} - \M A\,(\M C \Khat \M B)\Tra
  \bigr\rVert_F^2 ,
  \label{eq:A_unfolded}
\end{equation}
where %the iteration index is dropped for readability,
$(\T T + \bm{\Lambda})_{(1)} \in \R^{I \times JK}$ is the mode-1
unfolding and $\M C \Khat \M B \in \R^{JK \times R}$ the Khatri--Rao
product. Using the identity $(\M C \Khat \M B)\Tra (\M C \Khat \M B) = (\M C\Tra \M C) \Hada (\M B\Tra \M B)$, where $\Hada$ is the Hadamard product, the normal equations
of~\eqref{eq:A_unfolded} involve only
\begin{equation*}
  \M{G}_{\M A} = (\M C\Tra \M C) \Hada (\M B\Tra \M B) + \epsilon I ,
  \qquad
  \M{M}_{\M A} = (\T T + \bm{\Lambda})_{(1)}\,(\M C \Khat \M B),
\end{equation*}
a small $\epsilon I$ being added for numerical stability ($\epsilon = 10^{-6}$). %Let us illustrate the update on two cases. 
In the unconstrained case, $\mathcal{C}_A = \R^{I \times R}$, \eqref{eq:A_unfolded} is an ordinary least-squares problem, $\M A \;\leftarrow\; \M{M}_{\M A}\, \M{G}_{\M A}\Inv$. Another case we consider is the nonnegative case, $\mathcal{C}_A = \R_+^{I \times R}$, solved with standard nonnegative least-squares (NNLS) algorithms.

%\emph{(i) Unconstrained factors:} When $\mathcal{C}_A = \R^{I \times R}$, \eqref{eq:A_unfolded} is an ordinary least-squares problem: $\M A \;\leftarrow\; \M{M}_{\M A}\, \M{G}_{\M A}\Inv$. %with the closed-form solution
% \begin{equation}
%   \M A \;\leftarrow\; \M{M}_{\M A}\, \M{G}_{\M A}\Inv.
%   \label{eq:A_ls}
% \end{equation}
%\emph{(ii) Nonnegative factors:} When $\mathcal{C}_A = \R_+^{I \times R}$, no closed form exists. %but the problem is separable across the rows of $\M A$. %Expanding the squared norm in~\eqref{eq:A_unfolded} and discarding terms independent of $\M A$, 
% Each row $i$ of $\M A$ needs to solve the $R$-dimensional nonnegative quadratic program
% \begin{equation}
%   \M{A}(i,:)\Tra
%   = \argmin_{\V a \,\geq\, 0}\;
%   \V a\Tra \M{G}_{\M A}\, \V a \;-\; 2\,\V a\Tra \M{M}_{\M A}(i,:)\Tra ,
%   \label{eq:A_nnls}
% \end{equation}
% solved via a nonnegative least-squares (NNLS) routine. 

\noindent\textbf{Update of $\T T$.}
The $\T T$-update is the only step in which the nonlinearity $f$ appears. Both terms of~\eqref{eq:aug_lag} are separable across tensor entries, so it decomposes into $IJK$ scalar problems. Letting $x$, $t$, $\lambda$ and $\hat{x}$ denote the $(i,j,k)$-th entries of $\T X$, $\T T$, $\Lambda$ and $\That X = \llbracket \M A,\M B,\M C \rrbracket$, each entry is updated by solving
\begin{equation}
  t \;\leftarrow\; \argmin_{t}\;
  \bigl(x - f(t)\bigr)^2
  + \frac{\rho}{2}\,(t + \lambda - \hat{x})^2 ,
  \label{eq:t_scalar}
\end{equation}
with $\hat{x} = \sum_{r=1}^{R} a_{ir} b_{jr} c_{kr}$. For the MinMax nonlinearity under the Frobenius loss, \eqref{eq:t_scalar} is exactly the scalar subproblem arising in the matrix case and admits the same closed-form solutions derived in~\cite{awari2025alternating}; we refer the reader there for details. The $\T T$-update thus costs a single element-wise pass over the tensor. The implementation accompanying this paper also supports the three loss functions and four nonlinearities considered in \cite{awari2025alternating}, making in total 12 different models all handled by the same algorithm. %See \Cref{alg:admm_nlcp} for the implementation.

% To avoid this and obtain better, and identifiable, solutions, we add the following regularization to our model: 
% \[
% \mu \, \bigl\lVert c \mathcal E - \llbracket \bA,\bB,\bC \rrbracket
%   \bigr\rVert_F^2
% \]
% where $\mathcal E$ is the tensor of all ones, and 
% $c = a$ if $b=+\infty$, $c = b$ if $a = -\infty$, and 
% $c = \frac{a+b}{2}$ otherwise. This steers the solution towards the given interval, making the model prefer  approximation of  clipped entries closer to the given interval.  
% This will allow our model to work even if full slices are clipped. We used the scale-indenpendent regularization parameter $\mu = 10^{-6} \frac{\|\mathcal X\|_F^2}{\|\mathcal E\|_F^2} = 10^{-6} \frac{\|\mathcal X\|_F^2}{IJK}$ in the numerical experiments. 

% ══════════════════════════════════════════════════════════════════════════
\section{Numerical Experiments}
\label{sec:experiments}
% ══════════════════════════════════════════════════════════════════════════
%In this section we evaluate the three approaches in two settings. The first setting uses synthetically generated tensors, where clipping is applied symmetrically from both sides and the true factors are known. The second is a more realistic case based on simulated metabolomics data, where clipping is one-sided, mimicking an instrument detection limit that removes the smallest entries of the tensor. In both settings we ask how well each method recovers the underlying factors as the amount of clipping increases. 
In this section we consider two settings. The first setting uses synthetically generated tensors, where clipping is applied symmetrically from both sides and the true factors are known. The second is a more realistic case based on simulated metabolomics data, where clipping is one-sided, mimicking an instrument detection limit (LOD) that affects the smallest entries of the tensor. In both settings, we compare the following three methods in terms of how well they recover the underlying factors as the amount of clipping increases:

%\subsection{Models compared}
%\label{ssec:models}
% We compare three ways of handling the clipped entries depending on what the threshold means: treating the affected entries as missing and excluding them from the fit; setting them to the threshold and treating that as the true value; and treating the threshold as a bound on the value. The corresponding models are described below.
\begin{itemize}[leftmargin=*,nosep]
 \item \textbf{CP-missing}: the clipped entries are treated as missing
    and excluded from the fit, so that the model is estimated from the
    known entries alone. The threshold is discarded.
  \item \textbf{CP-imputation}: the clipped entries are set to the
    threshold and the CP model~\eqref{eq:cp} is fitted to the
    resulting complete tensor. The threshold is thus treated as the true
    value. 
  \item \textbf{NCP} (proposed): the clipped entries are set to the
    threshold as in CP-imputation, but the nonlinear
    model~\eqref{eq:nlcp} is fitted instead, with $f$ matching the
    clipping. %Thus, NCP treats the threshold as a bound. 
 %   We write NCP-MinMax for the two nonlinearity of~\Cref{ssec:nonlinear_cp}, and NN-MinMax \evrim{refer to as NCP-MinMax-NN?} when non-negativity is imposed on the factors. \ngc{Actually, we only use NCP everywhere, maybe simpler to use NCP? Otherwise, for consistency, we could also use CP-imputation-NN, etc. Hence I would remove this last sentence.}
\end{itemize}

\vspace{-0.5cm}
\paragraph*{A. Evaluation metric.} %\label{ssec:eval}

\noindent Factor recovery performance is assessed using the Factor Match Score (FMS), which compares factors (with normalized columns) of two CP models $\{\M{A}^{(1)},\M{B}^{(1)},\M{C}^{(1)}\}$ and
\{$\M{A}^{(2)},\M{B}^{(2)},\M{C}^{(2)}\}$ as \vspace{-0.2cm} 
\begin{equation*}
  \mathrm{FMS} = \frac{1}{R}\sum_{r=1}^{R}
  \bigl| {\V{a}_r^{(1)}}^\top {\V{a}_r^{(2)}} \bigr|\,
  \bigl| {\V{b}_r^{(1)}}^\top {\V{b}_r^{(2)}} \bigr|\,
  \bigl| {\V{c}_r^{(1)}}^\top {\V{c}_r^{(2)}} \bigr| ,
  \label{eq:fms} \vspace{-0.2cm}  
\end{equation*}
after finding the best matching permutation. $\mathrm{FMS} \in [0,1]$,
with $1$ indicating an exact match. 

%\noindent\textbf{Reproducibility} measures whether a method recovers thesame solution from different initializations, a prerequisite for interpreting the extracted factors~\cite{morup2025tensor,adali2022}. A run is counted as near-best if its final objective lies within $10^{-2}$ of the lowest value attained, and we report the mean pairwise FMS among these runs.
\vspace{-0.5cm}
\paragraph*{B. Implementation details.}
\label{ssec:implementation}

CP-imputation is fitted with \texttt{cp\_opt} from the Tensor Toolbox v3.7~\cite{acar2011cpopt}. CP-missing is fitted with \texttt{cp\_wopt}~\cite{acar2011} modified to include the ridge regularization term, $\frac{\gamma}{2}\bigl(\lVert \M A \rVert_F^2
  + \lVert \M B \rVert_F^2 + \lVert \M C \rVert_F^2\bigr)$, with $\gamma = 10^{-1}$. 
% Given a binary mask $\T W$ with $w_{ijk} = 1$ for
% observed entries and $w_{ijk} = 0$ for clipped ones, it solves 
% \begin{equation}
%   \min_{\M A, \M B, \M C} \;
%   \frac{1}{2}
%   \bigl\lVert \T W \Hada \bigl(\T X - \llbracket \M A, \M B, \M C
%   \rrbracket\bigr) \bigr\rVert_F^2
%   + \frac{\gamma}{2}\bigl(\lVert \M A \rVert_F^2
%   + \lVert \M B \rVert_F^2 + \lVert \M C \rVert_F^2\bigr) ,
%   \label{eq:cpwopt}
% \end{equation}
% where we use $\gamma = 10^{-1}$. 
Both CP models are solved using the limited memory BFGS-B algorithm. Lastly, NCP is solved with Algorithm~\ref{alg:admm_nlcp}. %the ADMM algorithm of~\Cref{ssec:admm}. 
For all models, we use $20$ random initializations and select the run with the lowest objective. All models were found to be reproducible at every clipping level in both experiments with a mean pairwise $\mathrm{FMS}\approx 1$ comparing the factors of near-best runs. %Convergence for \texttt{cp\_opt} and \texttt{cp\_wopt} is evaluated using a relative loss tolerance of $10^{-10}$ and a gradient norm tolerance of $10^{-5}$. For NCP, the algorithm terminates when the objective falls below $10^{-9}$ or when its change across 10 consecutive iterations is less than $10^{-10}$. 
All algorithms stopped due to either small absolute or relative change in the total loss. The code is available from \url{https://github.com/Atharva140298/ADMM-for-Nonlinear-CP.git}.
%All the codes and experiments are implemented in MATLAB\footnote{\url{https://anonymous.4open.science/r/ADMM-for-Nonlinear-CP-CE37/}}.

%\paragraph*{C. Data}
%\label{ssec:data}

%\noindent\textbf{Synthetic data:}

%\noindent\textbf{Metabolomics data}:

% ══════════════════════════════════════════════════════════════════════════

\vspace{-0.5cm}
\paragraph*{C. Results on Synthetic data.}
\label{ssec:results_synthetic}
% ══════════════════════════════════════════════════════════════════════════
%Since $\M A_0, \M B_0, \M C_0$ are known, 
%We generate rank-$R$ tensors $\T Y = \llbracket \M A_0, \M B_0, \M C_0 \rrbracket \in
%\R^{50 \times 40 \times 30}$ with $R = 3$ and factor matrix entries drawn randomly from a standard normal distribution, so that $\T Y$ contains both positive and negative values. Clipping is applied by shrinking the observation window symmetrically from both sides: for a clipping level $p\%$, the bounds are set to the $p/2$-th and $100-p/2$-th percentiles of $\T Y$, so that $p\%$ of the entries fall outside the window and are clipped, half at each end. We sweep $p$ from $0\%$ to $99\%$, which lets the two-sided MinMax nonlinearity be studied from the mildly clipped regime through to the point where almost nothing is observed. %To avoid conclusions that
%depend on a single draw
%We generate $10$ independent tensors and repeat the experiment on each.  
%FMS is computed between the factors recovered by each method from the clipped tensor $\T X$ and ground-truth factors $\M A_0, \M B_0, \M C_0$. No constraints are imposed on the factors when fitting the models. For each tensor the run attaining the lowest
%objective is retained. %and we report the mean and standard deviation of the resulting scores for the $10$ tensors. %The factors are unconstrained. 
We generate rank-$R$ tensors $\T Y = \llbracket \M A_0, \M B_0, \M C_0 \rrbracket \in \R^{50 \times 40 \times 30}$ with $R = 3$ and factor entries drawn from a standard normal distribution, so that $\T Y$ contains both
positive and negative values. Clipping shrinks the observation window symmetrically: at level $p\%$, the bounds are the $p/2$-th and $(100-p/2)$-th percentiles of $\T Y$, so that $p\%$ of the entries are clipped, half at each end. We sweep $p$ from $0\%$ to $99\%$, and repeat the experiment on $10$ independent tensors. FMS is computed against the
ground-truth factors $\M A_0, \M B_0, \M C_0$, retaining for each tensor the run with the lowest objective. No constraints are imposed on the factors.
\begin{figure}[htb]
\centering
\begin{tikzpicture}
\begin{axis}[
    %width=\columnwidth,
    %height=0.78\columnwidth,
    width=0.9\columnwidth,
height=0.624\columnwidth,
    xlabel={Clipped entries (\%)},
    ylabel={FMS vs.\ ground-truth},
    xmode=log,
    x dir=reverse,
    xmin=0.85, xmax=115,
    ymin=-0.06, ymax=1.10,
    xtick={100,50,20,10,5,2,1},
    xticklabels={0,50,80,90,95,98,99},
    minor xtick={},
    ytick={0,0.2,0.4,0.6,0.8,1.0},
    grid=major,
    grid style={gray!25, line width=0.3pt},
    legend pos=south west,
    legend style={font=\footnotesize, cells={anchor=west},
                  fill=white, draw=gray!50, row sep=-1pt},
    label style={font=\small},
    tick label style={font=\footnotesize},
    axis line style={gray!60},
    every axis plot/.append style={line width=1pt},
]

\addplot[black, dashed, line width=0.7pt, forget plot]
    coordinates {(0.85,1) (115,1)};

% =================== CP-missing ===================
\addplot[name path=misU, draw=none, forget plot] coordinates {
(100,1.000) (90,1.000) (80,1.000) (70,1.000) (60,1.000) (50,1.000)
(40,1.000) (30,0.997) (20,0.979) (10,0.743) (9,0.604) (8,0.361)
(7,0.058) (6,0.016) (5,0.003) (4,0.003) (3,0.003) (2,0.004) (1,0.004)};
\addplot[name path=misL, draw=none, forget plot] coordinates {
(100,1.000) (90,1.000) (80,1.000) (70,1.000) (60,1.000) (50,1.000)
(40,1.000) (30,0.997) (20,0.959) (10,0.383) (9,0.304) (8,0.041)
(7,0.000) (6,0.000) (5,0.003) (4,0.003) (3,0.003) (2,0.004) (1,0.004)};
\addplot[blue!70!black, opacity=0.18, forget plot]
    fill between[of=misU and misL];

\addplot[color=blue!70!black, mark=*, mark size=1.5pt,
         mark options={fill=blue!70!black}] coordinates {
(100,1.000) (90,1.000) (80,1.000) (70,1.000) (60,1.000) (50,1.000)
(40,1.000) (30,0.997) (20,0.969) (10,0.563) (9,0.454) (8,0.201)
(7,0.028) (6,0.006) (5,0.003) (4,0.003) (3,0.003) (2,0.004) (1,0.004)};
\addlegendentry{CP-missing}

% ================= CP-imputation =================
\addplot[name path=impU, draw=none, forget plot] coordinates {
(100,1.000) (90,0.961) (80,0.940) (70,0.913) (60,0.888) (50,0.865)
(40,0.843) (30,0.823) (20,0.805) (10,0.791) (9,0.790) (8,0.788)
(7,0.787) (6,0.792) (5,0.785) (4,0.784) (3,0.784) (2,0.828) (1,0.789)};
\addplot[name path=impL, draw=none, forget plot] coordinates {
(100,1.000) (90,0.961) (80,0.920) (70,0.893) (60,0.868) (50,0.845)
(40,0.823) (30,0.803) (20,0.785) (10,0.771) (9,0.770) (8,0.768)
(7,0.767) (6,0.752) (5,0.765) (4,0.764) (3,0.764) (2,0.668) (1,0.549)};
\addplot[green!45!black, opacity=0.18, forget plot]
    fill between[of=impU and impL];

\addplot[color=green!45!black, mark=triangle*, mark size=2.0pt,
         mark options={fill=green!45!black}] coordinates {
(100,1.000) (90,0.961) (80,0.930) (70,0.903) (60,0.878) (50,0.855)
(40,0.833) (30,0.813) (20,0.795) (10,0.781) (9,0.780) (8,0.778)
(7,0.777) (6,0.772) (5,0.775) (4,0.774) (3,0.774) (2,0.748) (1,0.669)};
\addlegendentry{CP-imputation}

% ===================== NCP =====================
\addplot[name path=ncpU, draw=none, forget plot] coordinates {
(100,1.000) (90,1.000) (80,1.000) (70,1.000) (60,1.000) (50,1.000)
(40,0.998) (30,0.990) (20,0.975) (10,0.909) (9,0.899) (8,0.887)
(7,0.875) (6,0.862) (5,0.849) (4,0.836) (3,0.833) (2,0.821) (1,0.811)};
\addplot[name path=ncpL, draw=none, forget plot] coordinates {
(100,1.000) (90,1.000) (80,1.000) (70,1.000) (60,1.000) (50,1.000)
(40,0.998) (30,0.990) (20,0.955) (10,0.889) (9,0.879) (8,0.867)
(7,0.855) (6,0.842) (5,0.829) (4,0.816) (3,0.793) (2,0.781) (1,0.771)};
\addplot[red!75!black, opacity=0.18, forget plot]
    fill between[of=ncpU and ncpL];

\addplot[color=red!75!black, mark=square*, mark size=1.5pt,
         mark options={fill=red!75!black}] coordinates {
(100,1.000) (90,1.000) (80,1.000) (70,1.000) (60,1.000) (50,1.000)
(40,0.998) (30,0.990) (20,0.965) (10,0.899) (9,0.889) (8,0.877)
(7,0.865) (6,0.852) (5,0.839) (4,0.826) (3,0.813) (2,0.801) (1,0.801)};
\addlegendentry{NCP}

\end{axis}
\end{tikzpicture}
\caption{Factor recovery on synthetic data as the observation window
shrinks symmetrically from both sides. %FMS is computed against the true
%factors; 
Solid lines show the mean FMS and shaded bands one standard deviation
over $10$ tensors. %, retaining the best of $20$ initializations per
%tensor.  
The dashed line marks perfect recovery.}
\label{fig:synth_sweep}
\end{figure}
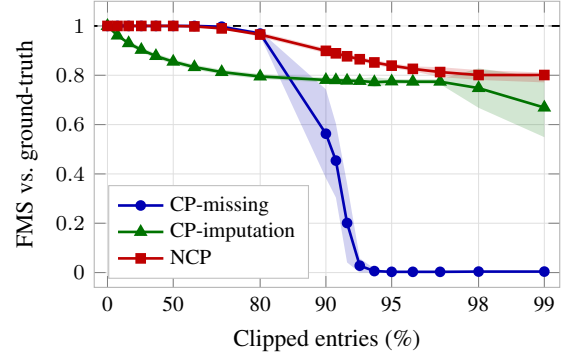
\Cref{fig:synth_sweep} reports factor recovery as the clipping becomes increasingly severe. CP-missing and NCP maintain high factor recovery up to $80\%$ clipping. In contrast, the FMS of CP-imputation decreases even under mild clipping and approaches a plateau near $0.78$.
%p to $80\%$ clipping, CP-missing and NCP both recover the true factors essentially exactly, while CP-imputation separates early and settles onto a plateau near $0.78$. 
The robustness of CP-missing to substantial clipping is expected~\cite{acar2011}. %A rank-$R$ CP
%model of an $I \times J \times K$ tensor has $R(I{+}J{+}K{-}1)+1$ free
%parameters. 
When the number of observed entries substantially exceeds the number of learned parameters, observed entries can provide sufficient information to identify the underlying low-rank structure~\cite{acar2011}. In contrast, CP-imputation treats clipping thresholds as exact measurements, introducing systematic errors that affect factor recovery even at low clipping levels. The persistence of this bias as clipping increases is consistent with the observed plateau in FMS.
%CP-imputation, by contrast, degrades from the outset and setting the clipped entries to the threshold asserts a value the data does not support. 
%The resulting bias does not disappear as more entries are clipped, which is why the curve flattens rather than falls.

Under severe clipping, structure of the remaining entries becomes critical. Beyond $90\%$ clipping, entire fibers and substantial portions of slices are clipped. %, producing structured regions of missing data. 
Such structured missingness poses greater challenges for factor recovery than randomly distributed missingness~\cite{acar2011}. Accordingly, the FMS of CP-missing decreases sharply %from $0.56$ 
at $90\%$. %to below $0.02$ at $93\%$. 
Increased variability across tensors indicates that recovery becomes less consistent. % before FMS approaches zero.

%The picture changes once clipping becomes severe. Beyond $90\%$, the
%clipped entries no longer scatter through the tensor but remove coherent
%regions of it, and whole fibers and large parts of slices are lost.
%Structured missingness of this kind is considerably harder than
%missingness spread at random~\cite{acar2011}. CP-missing collapses accordingly, falling from $0.56$ at $90\%$ to below $0.02$ by $93\%$, and the widening spread across tensors in that range shows the method becoming unpredictable before it fails outright.

In contrast, the FMS of NCP degrades gradually as clipping increases and remains higher than that of both CP models even under severe clipping. NCP retains an FMS of $0.80$ even when only $1\%$ of entries are observed, benefiting from the bound information provided by clipped entries.

%NCP degrades gracefully across the whole range and remains above both
%baselines throughout the tail, retaining an FMS of $0.80$ when only $1\%$ of the entries are observed. %Because the clipped entries are retained as bounds rather than discarded.

\noindent\textbf{Implicit regularization in the ADMM updates.}
When a full slice is clipped, the NCP solution need not be unique: any approximation of that slice lying beyond the threshold gives zero error, since the model maps it onto the threshold regardless. For $[a,b] = [1,+\infty]$ and a slice clipped to $a=1$, the all-zero slice is optimal, yet an approximation close to the threshold would be far more meaningful. Our ADMM algorithm favors the latter. It is initialized with $\T T = \T X$ and $\bm{\Lambda} = 0$, so the first iteration fits $\llbracket \M A, \M B, \M C \rrbracket$ to $\T X$ itself, clipped values included; it is effectively a CP-imputation step. Entries that land near the clipped values then stay there, since the subsequent updates leave $\T T$ close to those values and $\bm{\Lambda}$ close to zero. A further analysis of this is left as future work.%a extended version of this work. 
\vspace{-0.54cm}
\paragraph*{D. Results on Metabolomics data.}
\label{ssec:res_nmr}
% ══════════════════════════════════════════════════════════════════════════
We use the simulated data of Li et al.~\cite{li2024}, 
generated using a human metabolic model, %defined by ordinary differential equations, with metabolites as variables and kinetic constants as parameters, capturing multi-organ metabolic dynamics following a meal. Fifty virtual subjects were obtained by randomly perturbing the kinetic parameters to introduce individual variation, which yields cleaner dynamics than noisy real measurements while remaining physiologically realistic. The data contains six blood metabolites and hormones (insulin, glucose, pyruvate, lactate, alanine and $\beta$-hydroxybutyrate), measured at eight time points
forming a tensor $\T X \in \R^{50 \times 8 \times 6}$ with modes \emph{subjects} $\times$ \emph{time} $\times$ \emph{metabolites}. In real metabolomics measurements, concentrations below the instrument LOD are not reported~\cite{wei2018missing}. We emulate this detection limit by clipping  tensor entries from below. For a clipping level of $p\% \in [0,50\%]$, we set the threshold to the $p$-th percentile of $\T X$ and every entry below it is clipped. %so the affected entries are the $p\%$ lowest-abundance measurements. 
%We sweep $p$ from $0\%$ to $50\%$. 
Clipping does not spread evenly across the tensor, low-abundance metabolites lose a far larger fraction of their entries. Following~\cite{li2024}, the data is preprocessed after clipping by scaling each metabolite slice by its root-mean-square value on the observed entries. 

Unlike the synthetic experiments, ground-truth factors are not available for the simulated metabolomics data. We therefore use a nonnegative CP model fitted to the \emph{fully observed} tensor as the reference, denoted by CP-NN. This reference represents the solution obtained in the absence of detection limits and is available because the data is simulated. 
At each clipping level, FMS is computed between the factors estimated from the clipped tensor and those of CP-NN. The near-best CP-NN runs have a mean pairwise FMS of $1.00$, indicating that the reference factors are reproducible, a prerequisite for interpreting the extracted factors~\cite{morup2025tensor,adali2022}. 
%across random initializations. 
All models use $R=3$ components with nonnegativity constraints, consistent with the nonnegative concentrations. 

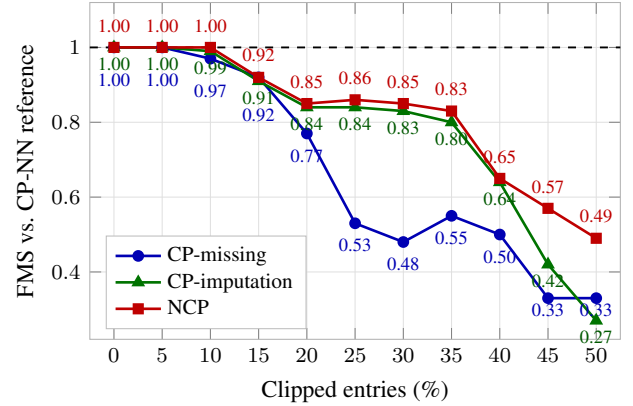
\begin{figure}[htb]
\centering
\begin{tikzpicture}
\begin{axis}[
    width=\columnwidth,
    height=0.702\columnwidth,
    xlabel={Clipped entries (\%)},
    ylabel={FMS vs.\ CP-NN reference},
    xmin=-2.5, xmax=52.5,
    ymin=0.22, ymax=1.12,
    xtick={0,5,10,15,20,25,30,35,40,45,50},
    ytick={0.2,0.4,0.6,0.8,1.0},
    grid=major,
    grid style={gray!25, line width=0.3pt},
    legend pos=south west,
    legend style={
        font=\footnotesize,
        cells={anchor=west},
        fill=white,
        draw=gray!50,
        row sep=-1pt,
    },
    label style={font=\small},
    tick label style={font=\footnotesize},
    axis line style={gray!60},
    every axis plot/.append style={line width=1pt},
    nodes near coords style={font=\scriptsize},
]

% reference line at perfect recovery
\addplot[black, dashed, line width=0.7pt, forget plot]
    coordinates {(-2.5,1) (52.5,1)};

% ==================================================
% CP-missing: curve and markers
% ==================================================
\addplot[
    color=blue!70!black,
    mark=*,
    mark size=1.7pt,
    mark options={fill=blue!70!black},
] coordinates {
    (0,1.00)
    (5,1.00)
    (10,0.97)
    (15,0.92)
    (20,0.77)
    (25,0.53)
    (30,0.48)
    (35,0.55)
    (40,0.50)
    (45,0.33)
    (50,0.33)
};
\addlegendentry{CP-missing}

% First 3 blue labels: yshift = -6pt
\addplot[
    draw=none,
    mark=none,
    forget plot,
    nodes near coords,
    point meta=explicit symbolic,
    every node near coord/.append style={
        below,
        yshift=-6pt,
        text=blue!70!black,
        font=\scriptsize
    },
] coordinates {
    (0,1.00)  [1.00]
    (5,1.00)  [1.00]
    (10,0.97) [0.97]
};

% Blue label 0.92: yshift = -8pt
\addplot[
    draw=none,
    mark=none,
    forget plot,
    nodes near coords,
    point meta=explicit symbolic,
    every node near coord/.append style={
        below,
        yshift=-8pt,
        text=blue!70!black,
        font=\scriptsize
    },
] coordinates {
    (15,0.92) [0.92]
};

% Intermediate blue labels: yshift = -2pt
\addplot[
    draw=none,
    mark=none,
    forget plot,
    nodes near coords,
    point meta=explicit symbolic,
    every node near coord/.append style={
        below,
        yshift=-2pt,
        text=blue!70!black,
        font=\scriptsize
    },
] coordinates {
    (20,0.77) [0.77]
    (25,0.53) [0.53]
    (30,0.48) [0.48]
    (35,0.55) [0.55]
    (40,0.50) [0.50]
};

% Last two blue labels (0.33): yshift = 2pt
\addplot[
    draw=none,
    mark=none,
    forget plot,
    nodes near coords,
    point meta=explicit symbolic,
    every node near coord/.append style={
        below,
        yshift=2pt,
        text=blue!70!black,
        font=\scriptsize
    },
] coordinates {
    (45,0.33) [0.33]
    (50,0.33) [0.33]
};

% ==================================================
% CP-imputation
% ==================================================
\addplot[
    color=green!45!black,
    mark=triangle*,
    mark size=2.0pt,
    mark options={fill=green!45!black},
    nodes near coords,
    point meta=explicit symbolic,
    every node near coord/.append style={
        below, yshift=0pt, text=green!35!black, font=\scriptsize},
] coordinates {
    (0,1.00)  [1.00]
    (5,1.00)  [1.00]
    (10,0.99) [0.99]
    (15,0.91) [0.91]
    (20,0.84) [0.84]
    (25,0.84) [0.84]
    (30,0.83) [0.83]
    (35,0.80) [0.80]
    (40,0.64) [0.64]
    (45,0.42) [0.42]
    (50,0.27) [0.27]
};
\addlegendentry{CP-imputation}

% ==================================================
% NCP
% ==================================================
\addplot[
    color=red!75!black,
    mark=square*,
    mark size=1.6pt,
    mark options={fill=red!75!black},
    nodes near coords,
    point meta=explicit symbolic,
    every node near coord/.append style={
        above, yshift=2pt, text=red!75!black, font=\scriptsize},
] coordinates {
    (0,1.00)  [1.00]
    (5,1.00)  [1.00]
    (10,1.00) [1.00]
    (15,0.92) [0.92]
    (20,0.85) [0.85]
    (25,0.86) [0.86]
    (30,0.85) [0.85]
    (35,0.83) [0.83]
    (40,0.65) [0.65]
    (45,0.57) [0.57]
    (50,0.49) [0.49]
};
\addlegendentry{NCP}

\end{axis}
\end{tikzpicture}
\caption{FMS between the factors recovered by the three methods on the clipped tensor and those recovered by CP-NN on the fully observed tensor. 
%The dashed line marks perfect recovery.
}
\label{fig:fms_sweep}
\end{figure}
%\vspace{3pt}
\Cref{fig:fms_sweep} reports factor recovery compared to the CP-NN reference across increasing clipping levels. All three methods maintain FMS values above $0.90$ up to $15\%$ clipping. %At higher clipping levels, 
FMS of CP-missing decreases sharply %from $0.77$ 
at $20\%$ clipping. % to $0.53$ at $25\%$. 
This decline reflects the structured information loss caused by clipping the lowest-abundance measurements first: at $20\%$ clipping, the entire slice of the lowest abundance metabolite falls below the detection limit. The detection limit thus induces structured missingness similar to that associated with reduced factor recovery in the synthetic experiments of \Cref{ssec:results_synthetic} \cite{acar2011}.
%Up to $15\%$ all three methods recover the reference factors almost perfectly. Beyond this point CP-missing breaks down sharply, falling from $0.77$ at $20\%$ to $0.53$ at $25\%$. The reason is structural because clipping removes the lowest-abundance measurements first. At $20\%$, the insulin slice falls entirely below the detection limit. This is again the structured missingness that hampered CP-missing in the synthetic experiments of \Cref{ssec:results_synthetic} \cite{acar2011}, here arising naturally from the detection limit.
CP-imputation and NCP maintain FMS values of at least $0.80$ up to $35\%$ clipping. The performance of CP-imputation over this range can be explained by the properties of the metabolomics data. The clipped entries correspond to sub-LOD concentrations of magnitude $10^{-3}$--$10^{-2}$ and lie close to the detection limit. Replacing them with this threshold therefore introduces only small perturbations, whose effect on the CP fit is comparable to that of low-level noise. Under more severe clipping, differences between the underlying concentrations and the threshold increase, making threshold imputation less representative of the original data. %At $45\%$ and $50\%$ clipping, NCP retains FMS values of $0.57$ and $0.49$, respectively, compared with $0.42$ and $0.27$ for CP-imputation. 
At the highest clipping level, CP-imputation factor recovery also falls below that of CP-missing. These results indicate that threshold imputation has a limited effect when the clipped concentrations lie close to the threshold, but increasingly distorts the estimated factors as these differences grow. NCP maintains FMS values at least as high as both baselines throughout the tested range by retaining the threshold as a valid bound, irrespective of its distance from the underlying concentrations.
%CP-imputation and NCP instead hold a plateau above $0.80$ until $35\%$. That CP-imputation does so well over this range is specific to the application. The clipped entries are sub-LOD concentrations, of magnitude $10^{-2}$--$10^{-3}$ at the clipping levels considered, so setting them to the detection limit perturbs the tensor only slightly, and the perturbation acts much like a small amount of noise, which a linear CP model handles well. The threshold is itself close to the clipped entries. This ceases to hold as clipping becomes more severe and the clipped values move further from the threshold. At $45\%$ and $50\%$, NCP retains $0.57$ and $0.49$ while CP-imputation falls to $0.42$ and $0.27$, dropping below even CP-missing at the highest level. Treating the threshold as a value is thus harmless while the clipped entries lie close to it, but becomes actively misleading once they do not. NCP outperforms both throughout, because the bound remains valid regardless of the clipped values.

Across synthetic and metabolomics experiments, increasing clipping affects the two baselines through different mechanisms. For CP-missing, factor recovery deteriorates when clipping removes most slice entries. For CP-imputation, factor recovery remains competitive only when the underlying values of the clipped entries lie close to the threshold.
%Taken together with the synthetic experiments, the two baselines fail in
%different scenarios and for different reasons, and neither failure is predictable from the clipping level alone: CP-missing collapses as soon as most of the  slice is lost, while CP-imputation is competitive only as long as the clipped values stay near the threshold.

% ══════════════════════════════════════════════════════════════════════════
\section{Conclusion} 
\label{sec:conclusions}
% ══════════════════════════════════════════════════════════════════════════
We introduced a nonlinear CP model together with a modular and
flexible ADMM algorithm for fitting it, and studied the pattern discovery performance. The setting we specifically
address is clipping, where each affected entry carries only the information that a
value lies beyond a known threshold. We compared two linear approaches for pattern discovery: treating clipped entries as missing %and excluding them from the fit, 
or imputing them at the threshold. Our experiments show distinct limitations of these approaches. %Excluding clipped entries impairs factor recovery when entire slices are lost, as occurs in the metabolomics experiment when a metabolite falls below the detection limit. Threshold imputation becomes less effective when the underlying values lie far from the threshold. 
Retaining the threshold as a bound allows NCP to use information that these linear approaches discard or misrepresent. NCP, therefore, performs as well as or better than these potential baselines. %Under severe clipping, NCP achieves higher FMS than both baselines, indicating better recovery of the true factors in the synthetic experiments and greater agreement with the CP-NN reference in the metabolomics experiment. These results support explicitly incorporating threshold information when clipping substantially affects the data.
%Retaining the threshold as a bound avoids both failure modes. Across the
%regimes considered, NCP recovered the underlying factors more accurately
%than either linear approach. Using the threshold explicitly is
%therefore worthwhile with the increasing severity of clipping or saturation.
%The proposed NCP approach is not specific to metabolomics and could also be applied in other applications with known measurement limits such as clipped audio processing.%. For example, in clipped audio, the recording limits provide bounds on the underlying signal~\cite{kitic2014audio}. 
We will explore other nonlinearities and loss functions for different applications as future work.

%Nothing in the formulation is specific to metabolomics: the same framework applies wherever measurements saturate, clipped audio~\cite{kitic2014audio} being the natural counterpart where the
%threshold is an upper bound on the recorded amplitude. Future work includes to use other nonlinearities and loss functions for other applications. 
\newpage 
\bibliographystyle{IEEEtran}
\bibliography{references}

\end{document}